\documentclass[aps,pra,twocolumn,showpacs,floatfix,twoside,superscriptaddress]{revtex4}
\usepackage{amsfonts}
\usepackage{amsmath}
\usepackage{amssymb}
\usepackage{graphicx}

\begin{document}

\title{Interference of Bose-Einstein condensates and entangled single-atom
state in a spin-dependent optical lattice}
\author{Linghua Wen}
\email{wenlinghua@wipm.ac.cn}
\author{Min Liu}
\author{Hongwei Xiong}
\email{xionghongwei@wipm.ac.cn}
\affiliation{State Key Laboratory
of Magnetic Resonance and Atomic and Molecular Physics, Wuhan
Institute of Physics and Mathematics, Chinese Academy of Sciences,
Wuhan 430071, P. R. China} \affiliation{Center for Cold Atom
Physics, Chinese Academy of Sciences, Wuhan 430071, P. R. China}
\affiliation{Graduate School, Chinese Academy of Sciences, Beijing
100080, P. R. China}
\author{Mingsheng Zhan}
\email{mszhan@wipm.ac.cn}
\affiliation{State Key Laboratory of Magnetic Resonance and Atomic and Molecular Physics,
Wuhan Institute of Physics and Mathematics, Chinese Academy of Sciences,
Wuhan 430071, P. R. China}
\affiliation{Center for Cold Atom Physics, Chinese Academy of Sciences, Wuhan 430071, P.
R. China}
\date{\today }

\begin{abstract}
We present a theoretical model to investigate the interference of
an array of Bose-Einstein condensates loaded in a one-dimensional
spin-dependent optical lattice, which is based on an assumption
that for the atoms in the entangled single-atom state between the
internal and the external degrees of freedom each atom interferes
only with itself. Our theoretical results agree well with the
interference patterns observed in a recent experiment by Mandel
\textit{et al}. [Phys. Rev. Lett. \textbf{91}, 010407 (2003)]. In
addition, an experimental suggestion of nonuniform phase
distribution is proposed to test further our theoretical model and
prediction. The present work shows that the entanglement of a
single atom is sufficient for the interference of the condensates
confined in a spin-dependent optical lattice and this interference
is irrelevant with the phases of individual condensates, i.e.,
this interference arises only between each condensate and itself
and there is no interference effect between two arbitrary
different condensates.
\end{abstract}

\pacs{03.75.Lm, 03.75.Gg, 05.60.Gg} \maketitle

\section{Introduction}

Since the first interference measurement \cite{R1} on two
expanding condensates there has been a growing interest in the
experimental and theoretical study on the interference of
Bose-Einstein condensates (BECs). In particular, optical lattices
created by retroreflected laser beams provide a unique tool for
testing at a fundamental level the quantum properties of BECs in a
periodic potential \cite{R2}. The interference patterns obtained
from the expansion of an array of condensates trapped in an
optical lattice are commonly used as a probe of the phase
properties of this system \cite{R3,R4,R5,R6}. Current
understanding of the interference of BECs is largely based on the
concept of phase coherence which reveals the superfluidity and the
matter wave nature of the condensates.

For the interference of two condensates, it is shown that an
interference pattern arises whether they are initially in phase
state (with locked relative phase) \cite{R1} or in Fock state
(with definite particle number) \cite{R7,R8,R9,R10}. In the latter
case, the interference effect is still originated from a
well-defined relative phase which is "built up" during the
sequences of measurement, i.e., the definite phase is derived from
the dynamic evolution of this system (initially with random
relative phase). For a fully coherent array of condensates in
optical lattices, the interference pattern obtained from the free
expansion is a natural result of the fixed relative phases between
different condensates belonging to consecutive wells
\cite{R3,R4,R5,R11,R12,R13,R14,R15,R16}. When
the coherent array of condensates enter the Mott insulating phase (MIP) \cite%
{R17,R18} in which phase coherence is lost, things become
complicated. The pioneering experiment \cite{R17} of the
superfluid to Mott-insulator transition demonstrated that when the
weakly interacting gas entered the MIP the interference pattern
became blurry even disappeared completely. Whereas for a strong
interacting gas it has been pointed out that a good measure for
this Mott transition was excitation spectrum rather than
interference pattern \cite{R19}. Besides, relevant theoretical
works \cite{R20,R21} based on a correlation function method imply
that even in the MIP interference pattern should also be observed
in a single measurement. On the other hand, a recent interference
experiment by Hadzibabic \textit{et al.} \cite{R22} states that
the periodicity of the optical lattice is sufficient for the
interference of an array of independent BECs even with no phase
coherence. A similar discussion is also found in a theoretical
Ref. \cite{R23}. It is therefore important to explore the physics
of interference patterns produced following the releasing and
expansion of BECs.

In this paper, we present a theoretical model to investigate the
interference of an array of BECs confined in a spin-dependent
optical lattice, which is motivated by a recent experiment by
Mandel \textit{et al.} \cite{R24}. The interference patterns
obtained from our theoretical model and numerical calculations
agree well with those observed in the experiment \cite{R24}. Our
conclusion is that the interference of an array of Bose
condensates trapped in a spin-dependent optical lattice results
from the entanglement of a single atom and the interference is
irrelevant with the phases of individual condensates.

This paper is organized as follows: In Sec.~2, after introducing
the basic spirit of the experiment \cite{R24} a theoretical model
is presented. In Sec.~3, we calculate numerically the density
distributions of the wave packets in spin states $\left\vert
1\right\rangle $ and $\left\vert 0\right\rangle $, respectively.
Then we compare the theoretical interference patterns with those
observed in the experiment. The following Sec.~4 deals with an
experimental suggestion of nonuniform phase distribution, which
can be used to test further the theoretical model and prediction.
Finally, discussion and conclusion is given in Sec.~5.

\section{Theoretical model}

For a Bose-condensed gas in a harmonic magnetic trap and a
three-dimensional (3D) optical lattice, the atoms are localized on
individual lattice sites when the system enters the MIP. After
switching off the magnetic trap and the lattice potentials along
the $y$ and $z$ directions there only exists a 1D spin-dependent
optical lattice along the $x$ direction which is formed by two
counterpropagating laser beams with linear polarization vectors
enclosing an angle $\theta $. Then each atom is prepared into a
coherent superposition of two spin states $\left\vert
0\right\rangle \equiv \left\vert F=1,m_{F}=-1\right\rangle $ and
$\left\vert 1\right\rangle \equiv \left\vert
F=2,m_{F}=-2\right\rangle $ using a microwave pulse. By changing
the polarization angle $\theta $ one can realize the splitting and
transport of atomic wave packets such that the
wave packets of an atom in spin states $\left\vert 0\right\rangle $ and $%
\left\vert 1\right\rangle $ respectively are transported in
opposite directions, which is the so-called spin-dependent
transport. Finally, the optical lattice is turned off and the
emerging interference patterns can be used as a diagnosis signal
for the coherence of the spin-dependent transport. In a word, the
spin-dependent transport is the principal idea of the experiment
by Mandel \textit{et al. }\cite{R24}.

Our starting point is that there is an array of Bose condensates
formed in the 1D spin-dependent optical lattice along the
horizontal $x$ direction when the magnetic trap and the lattice
potentials along the $y$ and $z$ directions are turned off. Since
the system experienced a Mott transition in advance these
condensates do not have any phase coherence relative to each other
any more. In this situation, the tunnelling between neighboring
lattice sites is suppressed and the effects of the atomic
interactions during the expansion of the condensates can be
neglected, which holds in the experiment. Since each condensate
confined in the lattice potential is fully coherent, in the frame
of single particle theory it can be described by a single order
parameter $\Psi _{k}=(N/(2k_{M}+1))^{1/2}\exp [i\theta _{k}]\phi
_{k}$ according to the Hartree or mean-field approximation
\cite{R25}, where $\phi _{k}$ is the single-particle wave function
in the $k$th lattice site and $\theta _{k}$ is the initially
random relative phase of the $k$th condensate. The coefficient
$N/(2k_{M}+1)$ represents the average particle number of each
condensate, with $N$ denoting the total particle number of the
whole condensate, and $2k_{M}+1$ being the total number of lattice
sites.

Following the experimental manipulation sequence in \cite{R24}, we
now
consider an atom with two spin states $\left\vert 0\right\rangle $ and $%
\left\vert 1\right\rangle $ forming its two logical basis-vectors.
Initially, the atom lies in spin state $\left\vert 0\right\rangle
_{k}$. Without loss of generality, by using an initial arbitrary
$\alpha $ microwave pulse to drive Rabi oscillations between the
two spin states, the atom can be placed into a coherent
superposition of the two spin states $\left\vert 0\right\rangle
_{k} $ and $\left\vert 1\right\rangle _{k}$,

\begin{equation}  \label{evolution rule}
\left\{%
\begin{array}{ll}
\left\vert 0\right\rangle _{k}\to\cos [\frac{\alpha
}{2}]\left\vert 0\right\rangle _{k}+i\sin [\frac{\alpha
}{2}]\left\vert 1\right\rangle _{k},
&  \\
\left\vert 1\right\rangle _{k}\to i\sin [\frac{\alpha
}{2}]\left\vert 0\right\rangle _{k}+\cos [\frac{\alpha
}{2}]\left\vert 1\right\rangle _{k}.
&  \\
\end{array}%
\right.
\end{equation}
After a spin-dependent transport, the spin state\ of the atom is given by $%
\cos [\alpha /2]\left\vert 0\right\rangle _{k}+i\exp [i\beta ]\sin
[\alpha /2]\left\vert 1\right\rangle _{k+r}$, where the spatial
wave packet of the atom is split into two components in states
$\left\vert 0\right\rangle _{k}$ and $\left\vert 1\right\rangle
_{k+r}$, respectively, i.e., the atomic wave packet is delocalized
over the $k$th and the $(k+r)$th lattice site. In the above
notation, the wave packet in state $\left\vert 0\right\rangle $
has retained the original lattice site index. Here, $r$ denotes
the separation between two wave packets which are originated from
the same $k$th lattice site. The relative phase $\beta $ between
the two wave packets, being independent of the number of
particles, is determined by the accumulated kinetic and potential
energy phase in the transport process. With the choice of
parameters in the experiment, the phase $\beta $ is almost
constant throughout the cloud of atoms and its absolute value is
small. Consequently, the atomic wave function can be described by
an entangled single-atom state, i.e., an entangled quantum state
between the internal degree of freedom (spin) and the external
degree of freedom (spatial wave packet)

\begin{equation}  \label{entangled state}
\psi _{k}=\cos [\frac{\alpha }{2}]\left\vert 0\right\rangle
_{k}\varphi _{k}+i\exp [i\beta ]\sin [\frac{\alpha }{2}]\left\vert
1\right\rangle _{k+r}\varphi _{k+r}.
\end{equation}

We assume that the spatial wave packet has a form of Gaussian
distribution in coordinate space, i.e., $\varphi _{k}=A\exp
[-(x-kd)^{2}/2\sigma ^{2}]$, where $d=\lambda /2$ is the period of
the optical lattice and $\lambda $ is the wavelength of the
retroreflected laser beams. $A=1/\sigma ^{1/2}\pi ^{1/4}$ is a
normalization constant, and $\sigma $ denotes the width of the
condensate in each optical well. By applying a final $\pi /2$
microwave pulse whose transform rule is given by
Eq.(\ref{evolution rule}), one has

\begin{equation}
\phi _{k}=\left\vert 0\right\rangle \Xi _{0,k}+\left\vert
1\right\rangle \Xi _{1,k},  \label{erasing which way information}
\end{equation}%
where $\Xi _{0,k}$ and $\Xi _{1,k}$ are respectively given by

\begin{eqnarray}
\Xi _{0,k} &=&\frac{A}{\sqrt{2}}\{\cos [\frac{\alpha }{2}]\exp [-\frac{%
(x-kd)^{2}}{2\sigma ^{2}}]  \nonumber \\
&&-\sin [\frac{\alpha }{2}]\exp [i\beta -\frac{(x-(k+r)d)^{2}}{2\sigma ^{2}}%
]\},  \label{wave function of spin 0} \\
\Xi _{1,k} &=&\frac{iA}{\sqrt{2}}\{\cos [\frac{\alpha }{2}]\exp [-\frac{%
(x-kd)^{2}}{2\sigma ^{2}}]  \nonumber \\
&&+\sin [\frac{\alpha }{2}]\exp [i\beta -\frac{(x-(k+r)d)^{2}}{2\sigma ^{2}}%
]\}.  \label{wave function of spin 1}
\end{eqnarray}%
In Eq.(\ref{erasing which way information}), the indices of spin states $%
\left\vert 0\right\rangle $ and $\left\vert 1\right\rangle $ are
removed in view of the bosonic identity.

Once the spin-dependent optical lattice is switched off, the
evolution of the spatial components $\Xi _{j,k}(x,t)$ $(j=0,1)$ of
the atomic wave function can be derived by the propagator method
\cite{R15,R16,R25}

\begin{equation}
\Xi _{j,k}(x,t)=\int\nolimits_{-\infty }^{\infty }K(x,t;y,t=0)\Xi
_{j,k}(y,t=0)dy,  \label{integral equation}
\end{equation}%
where $\Xi _{j,k}(y,t=0)$ $(j=0,1)$ are the spatial components at
the initial time $t=0$ which are given by Eqs.(\ref{wave function
of spin 0}) and (\ref{wave function of spin 1}), and
$K(x,t;y,t=0)$ is the propagator in free space expressed as
\cite{R26}

\begin{equation}
K(x,t;y,t=0)=[\frac{m}{2\pi i\hbar t}]^{\frac{1}{2}}\exp [\frac{im}{2\hbar t}%
(x-y)^{2}].  \label{propagator}
\end{equation}%
By combining the formulae (\ref{wave function of spin 0})-(\ref{propagator}%
), one can obtain the following analytical results of the spatial
components after a straightforward calculation:

\begin{eqnarray}
\Xi _{0,k}(x,t) &=&\frac{A}{\sqrt{2(1+i\gamma t)}}\{\cos [\frac{\alpha }{2}%
]\exp [-\frac{(x-kd)^{2}}{2\sigma ^{2}(1+i\gamma t)}]  \nonumber \\
&&-\sin [\frac{\alpha }{2}]\exp [i\beta
-\frac{(x-(k+r)d)^{2}}{2\sigma
^{2}(1+i\gamma t)}]\},  \label{analytical result 0} \\
\Xi _{1,k}(x,t) &=&\frac{iA}{\sqrt{2(1+i\gamma t)}}\{\cos [\frac{\alpha }{2}%
]\exp [-\frac{(x-kd)^{2}}{2\sigma ^{2}(1+i\gamma t)}]  \nonumber \\
&&+\sin [\frac{\alpha }{2}]\exp [i\beta
-\frac{(x-(k+r)d)^{2}}{2\sigma ^{2}(1+i\gamma t)}]\},
\label{analytical result 1}
\end{eqnarray}%
where the parameter $\gamma =\hbar /m\sigma ^{2}$ denotes the
trapping frequency within a single well of the optical lattice.

We now consider the density distribution of the overall
condensates after switching off the spin-dependent optical
lattice. The wave function of the whole sample at time $t$ can be
expressed as

\begin{equation}
\Psi (x,t)=\sum_{k=-k_{M}}^{k_{M}}\sqrt{\frac{N}{2k_{M}+1}}\exp
[i\theta _{k}]\phi _{k}(x,t),  \label{total wave function}
\end{equation}%
where the time-dependent atomic wave function is given by $\phi
_{k}(x,t)=\left\vert 0\right\rangle \Xi _{0,k}(x,t)+\left\vert
1\right\rangle \Xi _{1,k}(x,t)$, and $\theta _{k}$ denotes the
random phase of the $k$th condensate at time $t$. In
Eq.(\ref{total wave function}), we have neglected the phase
diffusion of each condensate possibly induced by quantum and
thermal fluctuations, which won't affect the essential of our
present problem.

For a Bose-condensed gas confined in a trap, the phase fluctuation
is
characterized by the fluctuations in the chemical potential \cite%
{R27,R28,R29}. In the presence of a 1D optical lattice with
sufficiently strong intensity, the phase fluctuations for
different condensates in individual wells are independent from
each other. Two dominating physical ingredients are responsible
for the creation of phase diffusion: one is the collision between
condensed atoms and background hot cloud (thermal fluctuation),
and the second is spontaneously collective excitation due to
quantum fluctuation \cite{R29}. In real experiments, the phase
diffusion effect is small and can be omitted safely. Actually,
when taking into account the phase diffusion effect, the holistic
characters of interference pattern will not change except that the
central peak of the interference pattern will decrease a little
relatively \cite{R29}.

Obviously, there is no interference between the wave packets in
different spin states as the two logical basis-vectors $\left\vert
0\right\rangle $ and $\left\vert 1\right\rangle $\ are orthogonal.
The following model is based on an assumption that for the atoms
in the entangled single-atom state between the internal and
external degrees of freedom each atom interferes only with itself.
Concretely, the density distributions of the wave packets in spin
states $\left\vert 0\right\rangle $ or $\left\vert 1\right\rangle
$, i.e., the density distributions of the overall condensates
confined in the total occupied lattice sites, are not expressed by
Eqs.(\ref{density of spin 0}) and (\ref{density of spin 1})

\begin{eqnarray}
n_{0}(x,t) &=&\left\vert \sum_{k=-k_{M}}^{k_{M}}\sqrt{\frac{N}{2k_{M}+1}}%
\exp [i\theta _{k}]\Xi _{0,k}(x,t)\right\vert ^{2},
\label{density of spin 0} \\
n_{1}(x,t) &=&\left\vert \sum_{k=-k_{M}}^{k_{M}}\sqrt{\frac{N}{2k_{M}+1}}%
\exp [i\theta _{k}]\Xi _{1,k}(x,t)\right\vert ^{2}, \label{density
of spin 1}
\end{eqnarray}%
but given by Eqs.(\ref{simplified density of spin 0}) and
(\ref{simplified density of spin 1})

\begin{eqnarray}
n_{0}(x,t) &=&\frac{N}{2k_{M}+1}\sum_{k=-k_{M}}^{k_{M}}\left\vert
\Xi
_{0,k}(x,t)\right\vert ^{2},  \label{simplified density of spin 0} \\
n_{1}(x,t) &=&\frac{N}{2k_{M}+1}\sum_{k=-k_{M}}^{k_{M}}\left\vert
\Xi _{1,k}(x,t)\right\vert ^{2}.  \label{simplified density of
spin 1}
\end{eqnarray}

The test criterion of this model depends on whether its
theoretical prediction accords with the experimental results,
i.e., whether this model can interpret well the experiment. Then
we perform a Monte-Carlo analysis of $n_{1}(x,t)$ by assigning
sets of random numbers to the phase $\{\theta _{k}\}$. Our
simulation results show that the interference patterns based on
Eq.(\ref{density of spin 1}) don't agree with those observed in
the experiment \cite{R24} at all. In addition, provide that there
are locked phases for individual condensates, i.e., the phase
$\theta _{k}$ is not random (a simplest case is that the phase of
each condensate is the same), the calculation also shows that the
interference patterns derived from Eq.(\ref{density of spin 1})
are not in agreement with the experimental results. Thus it is
implied that Eqs.(\ref{density of spin 0}) and (\ref{density of
spin 1}) are invalid in explaining the experiment. As expected,
however, we find that the theoretical interference patterns based
on Eq.(\ref{simplified density of spin 1}) agree well with the
observed interference patterns in the experiment (see Fig.1-Fig.2
in section 3), which indicates that this model, i.e.
Eqs.(\ref{simplified density of spin 0}) and (\ref{simplified
density of spin 1}), can be employed to describe the real physics
of the emerging interference patterns in the experiment
\cite{R24}.

The physical essence of the density distributions expressed by Eqs.(\ref%
{simplified density of spin 0}) and (\ref{simplified density of
spin 1}) is that due to the entanglement of a single atom each
atom interferes only with itself (or each condensate interferes
only with itself), i.e., there is no interference effect between
two arbitrary different condensates. In other words, the
entanglement of a single atom is sufficient for the interference
of the overall condensates and this interference will be
irrelevant with the phases of individual condensates.

\section{Density distributions and evolution}

By using the experimental parameters in \cite{R24}, we plot the
density distributions of the atomic wave packets in states
$\left\vert
0\right\rangle $ and $\left\vert 1\right\rangle $ respectively based on Eqs.(%
\ref{simplified density of spin 0}) and (\ref{simplified density
of spin 1}).

\subsection{Parameters}

In the following calculations, the relevant parameters are
consistent with
those in the experiment, where $\alpha =\pi /2$, $N=3\times 10^{5}$, $%
\lambda =785$ nm, and $d=392.5$ nm. For simplicity, we treat the
relative phase $\beta $ as zero. Nevertheless, we'll also take
into account the effect of it on the interference patterns to
compare with the omitted case. Since the value of $\sigma $, which
characterizes the width of condensate in
each lattice site, is chiefly determined by the optical confinement \cite%
{R11}, one can evaluate it in terms of a variational calculation.
As a result, the ratio $\sigma /d=0.173$ is obtained. The total
number of the lattice sites $2k_{M}+1$ can be determined theoretically by the formula  $%
k_{M}^{2}=2\hbar \varpi (15Nad/8\pi ^{1/2}a_{ho}\sigma
)^{2/5}/(m\omega _{x}^{2}d^{2})$ (see Eq.(10) in \cite{R11}),
where the geometric average of the magnetic frequencies $\varpi
_{x}=2\pi \times 16$ Hz, $m$ is the mass of
$^{87}$Rb atom, the oscillator atom, the oscillator length $a_{ho}=\sqrt{%
\hbar /m\varpi }$, and the s-wave scattering length for $^{87}$Rb atom is $%
a\sim 50$ \r{A}. Thus $k_{M}\sim 50$ is obtained from the above
equation.

\subsection{Density distributions in state $\left\vert 1\right\rangle $}

In order to compare with the experiment, we consider firstly the
density distribution of the wave packets in state $\left\vert
1\right\rangle $ after the optical lattice is switched off with a
time of flight being 14 ms. The
analytical result of $n_{1}(x,t)$ at time $t$ is given by Eq.(\ref%
{simplified density of spin 1}). Shown in Fig.1(a) is the density
distribution (in units of $H=NA^{2}/(2k_{M}+1)$) in state
$\left\vert 1\right\rangle $ at $t=14$ ms after initially
localized atoms have been delocalized over two lattice sites. Note
that in all the figures plotted in this paper the horizontal
coordinate $x$ is in units of $\mu $m and the vertical coordinate
is in units of $H=NA^{2}/(2k_{M}+1)$. The density distributions in
the cases that initially localized atoms have been delocalized
over three (b), four (c), five (d), six (e), and seven (f) lattice
sites are given in Fig.1(b)--(f), respectively, where the
delocalized extension\ is denoted by $r$. With the separation $r$
increasing, we see that the fringe spacing of interference
patterns decreases remarkably and the visibility of the
interference patterns reduces distinctively (see Fig.1), which is
in agreement with the experimental results (see Fig.4 in
\cite{R24}).

\begin{figure}[h]
\centerline{\includegraphics[width=6cm,angle=-90]{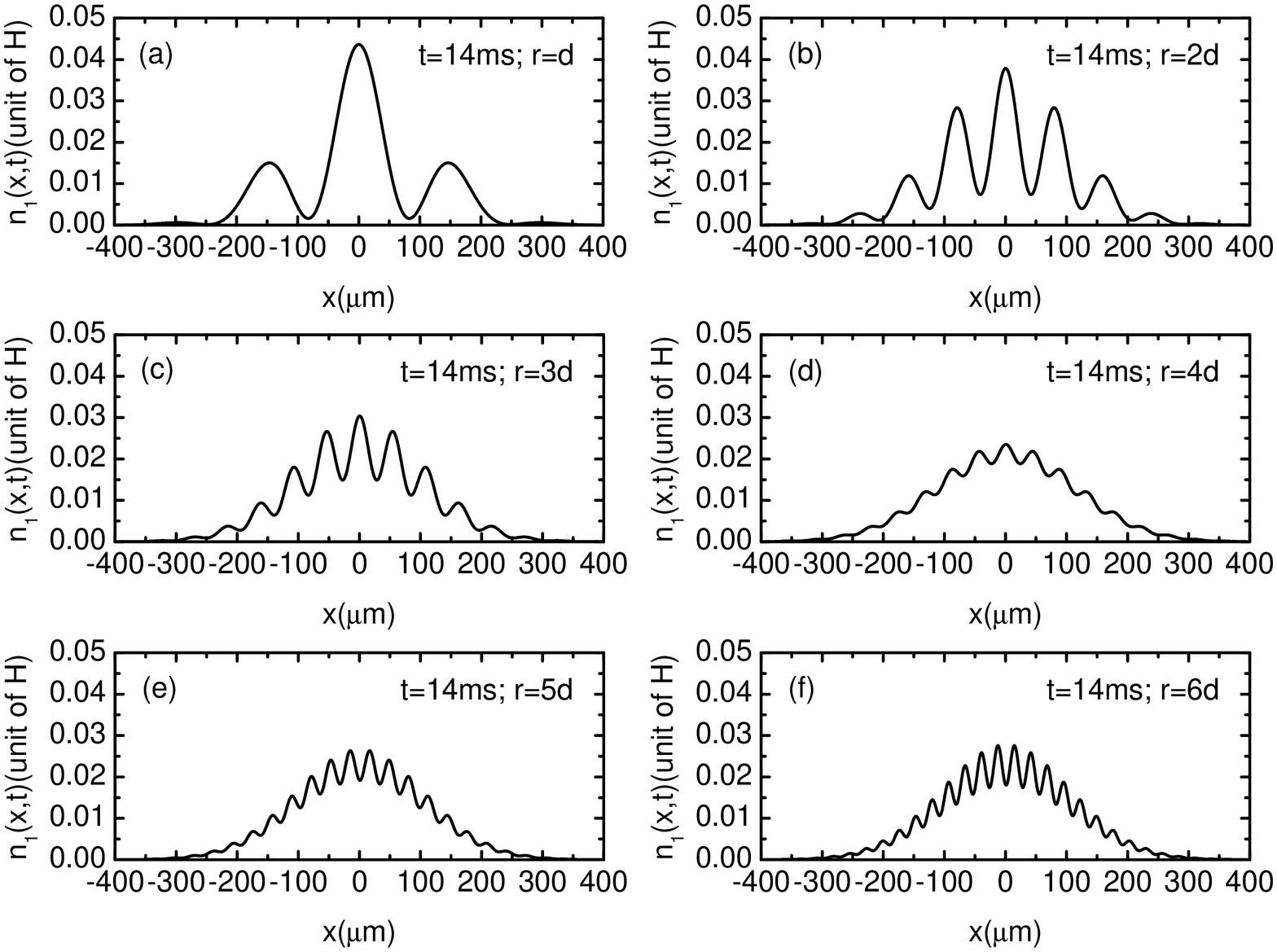}}
\caption{Density distributions in state $\left\vert 1\right\rangle
$ after switching off the spin-dependent optical lattice in the
cases that initially localized atoms have been delocalized over
two (a), three (b), four (c), five (d), six (e), and seven (f)
lattice sites by the interferometer sequence (see Fig.3 in
\protect\cite{R24}). The time of flight period is $14$
ms. The vertical coordinate $n_{1}(x,t)$ is in units of $H$ ($%
H=NA^{2}/(2k_{M}+1)$) and the horizontal coordinate $x$ is in units of $%
\protect\mu $m. $r$\ denotes the separation between the two wave
packets originated from the same lattice site.}
\end{figure}

In Fig.2, we show the evolution of the density distribution in state $%
\left\vert 1\right\rangle $ after initially localized atoms have
been delocalized over three lattice sites. Displayed in Fig.2(b)
is the density distribution at $t=15$ ms, which agrees well with
the observed interference pattern in the experiment (see Fig.5 in
\cite{R24}).

\begin{figure}[h]
\centerline{\includegraphics[width=2cm,angle=-90]{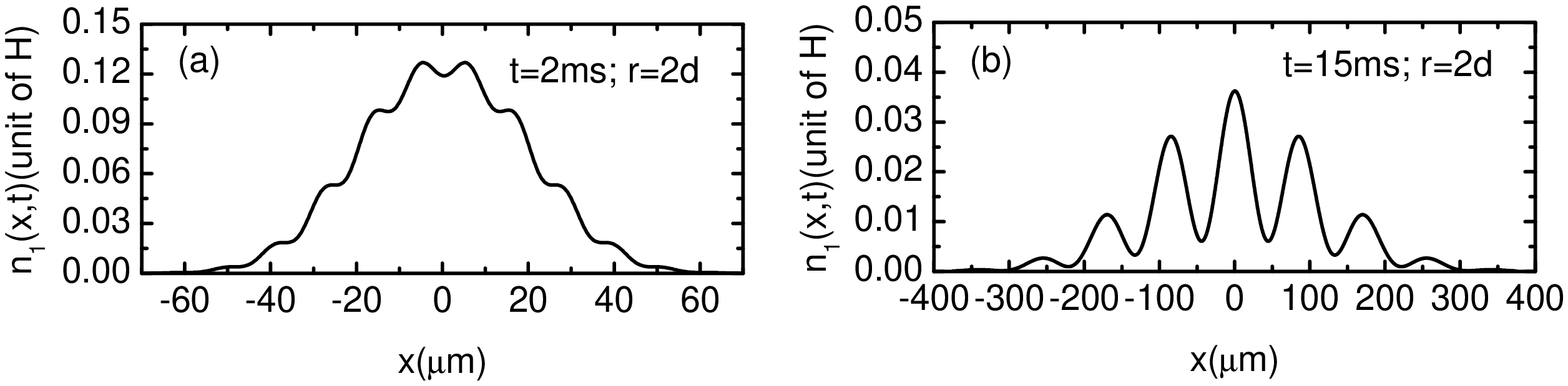}}
\caption{Evolution of the density distribution in state
$\left\vert 1\right\rangle $ with time $t$ after switching off the
spin-dependent optical lattice in the case that initially
localized atoms have been delocalized over three lattice sites.
The density distributions are shown at $t=2$ ms (a) and $t=15$ ms
(b).}
\end{figure}

In the calculations mentioned above, we have neglected the effect
of the relative phase $\beta $ on the density distribution by
treating it as zero.
Displayed in Fig.3 are the density distributions for the relative phase $%
\beta $ with $-\pi /12$ and $-\pi /3$ respectively. When taking
into account the relative phase $\beta $ between the two wave
packets the right-hand side peaks of the density distributions
become higher than the left-hand side ones, which breaks the
symmetry of the interference patterns to a certain extent. In
addition, the larger the absolute value of the phase $\beta $ is,
the weaker the symmetry of the interference pattern becomes.
According to the observed interference patterns, we can conclude
that the absolute value of the phase $\beta $ is possibly close to
zero in the experiment \cite{R24}.

\begin{figure}[h]
\centerline{\includegraphics[width=2cm,angle=-90]{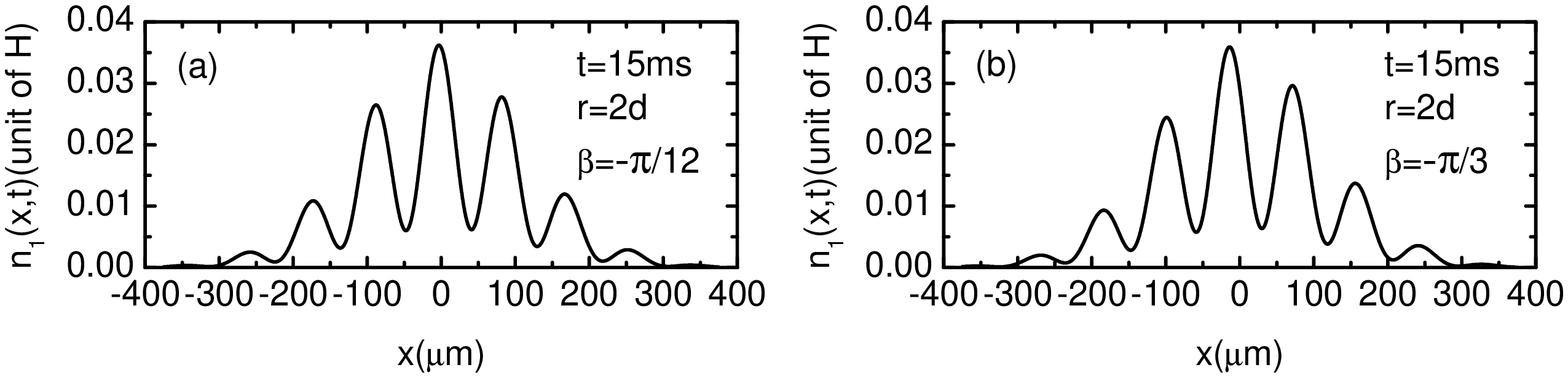}}
\caption{The effect of the phase $\protect\beta $ on the density
distribution in state $\left\vert 1\right\rangle $ after switching
off the spin-dependent optical lattice in the case that initially
localized atoms have been delocalized over three lattice sites.
The time of flight period is
$15$ ms. The density distributions are shown at $\protect\beta =-\protect\pi %
/12$ (a) and $\protect\beta =-\protect\pi /3$ (b), respectively.}
\end{figure}

\subsection{Density distributions in state $\left\vert 0\right\rangle $}

Now, we discuss the density distributions in state $\left\vert
0\right\rangle $ which were not observed in \cite{R24}. Similarly
to the foregoing analysis, the analytical result of the density
distributions in state $\left\vert 0\right\rangle $ is given by
Eq.(\ref{simplified density
of spin 0}). Shown in Fig.4 are the density distributions in state $%
\left\vert 0\right\rangle $ at time $t=14$ms. In contrast with the
density distributions in state $\left\vert 1\right\rangle $, the
positions of the sharp peaks in Fig.4(a)--(f) just become those of
local minimum densities in Fig.1(a)--(f) and vice versa, which can
be interpreted by the conservation of energy and particle number.

\begin{figure}[h]
\centerline{\includegraphics[width=6cm,angle=-90]{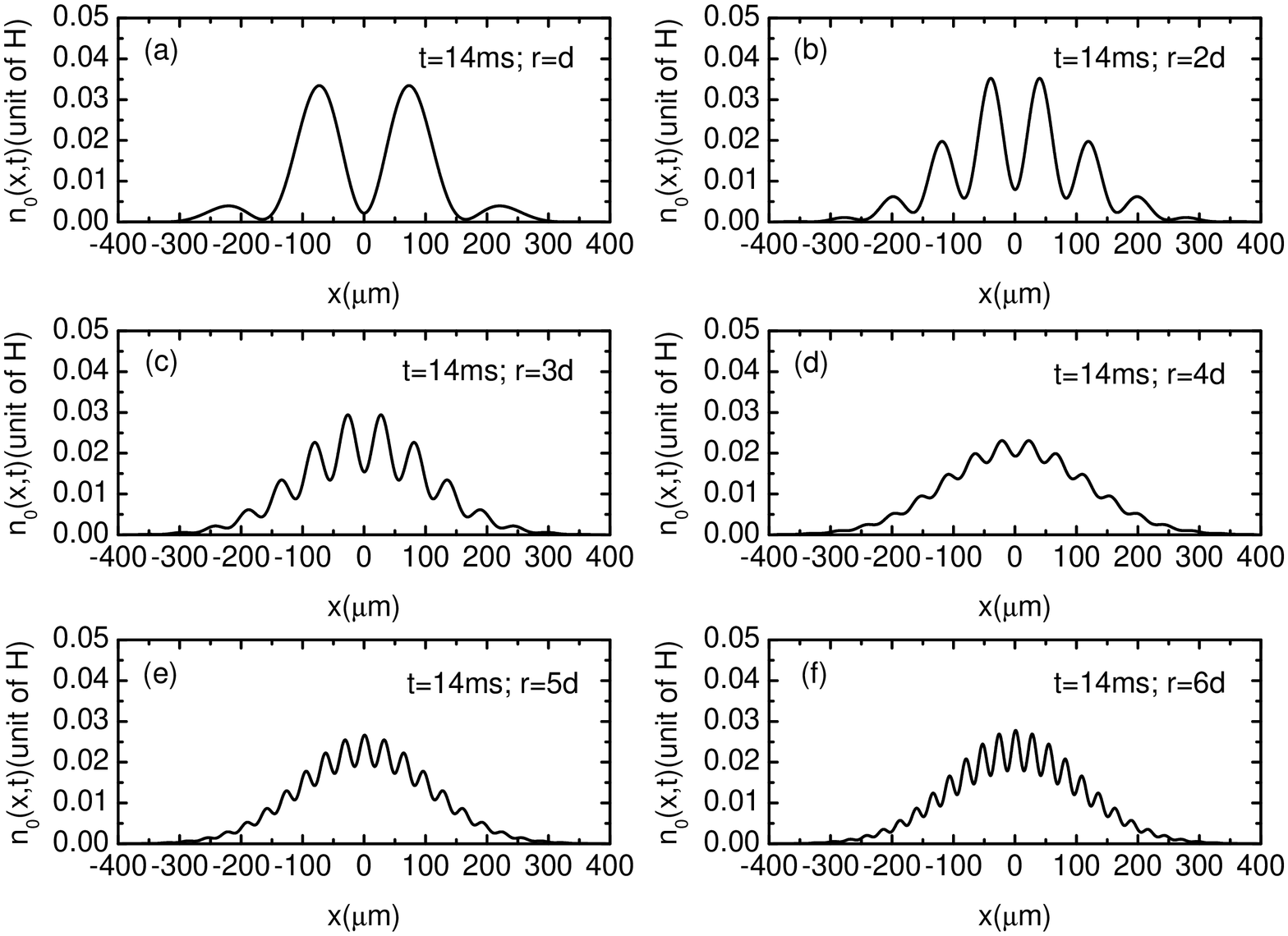}}
\caption{Density distributions in state $\left\vert 0\right\rangle
$ after switching off the spin-dependent optical lattice in the
cases that initially localized atoms have been delocalized over
two (a), three (b), four (c), five (d), six (e), and seven (f)
lattice sites. The time of flight period is $14$ ms.}
\end{figure}

\section{Experimental suggestion}

As mentioned above, the Bose condensates confined in the 1D
spin-dependent optical lattice have no phase coherence relative to
each other as the system experienced a Mott transition beforehand.
In this situation, the density distributions of the atomic wave
packets based on our theoretical model are in good agreement with
the observed interference patterns in \cite{R24}. From the
theoretical model (see Eqs.(\ref{simplified density of spin 0})
and (\ref{simplified density of spin 1})), the density
distributions of the atomic wave packets in different spin states
are irrelevant with the phases of individual condensates in this
system. Hence, when there is a locked phase distribution for the
array of condensates, the density distributions will not change.

To test further the validity of this model, we propose an
experimental suggestion of nonuniform phase distribution.
Concretely, we design a fixed linear phase distribution with a
total width of $2\pi $ for the array of condensates, in which the
phase difference between two neighboring condensates is $\delta
\theta =2\pi /(2k_{M}+1)$ ($k_{M}\sim 50$), i.e., the phase of the
$k$th condensate can
be expressed by $\theta _{k}=2\pi (k_{M}+k)/(2k_{M}+1)$ ($k=-k_{M},...,k_{M}$%
). This goal can be achieved by using the techniques of phase
redistributions such as phase imprinting \cite{R30,R31} and phase
engineering \cite{R32,R33}. Once a nonuniform phase distribution
is
performed successfully on the array of condensates, one applies an initial $%
\alpha =\pi /2$ microwave pulse to drive Rabi oscillations between
the two spin states $\left\vert 0\right\rangle $ and $\left\vert
1\right\rangle $, respectively. Thus all the atoms initially in
spin state $\left\vert 0\right\rangle $ are placed in a coherent
superposition of the two spin states, where the transform rule is
given by Eq.(\ref{evolution rule}).

The following deduction is similar to that in the preceding
sections. After a spin-dependent transport and applying a final
$\pi /2$ microwave pulse as well as a releasing of the optical
lattice, the density distributions of the wave packets in states
$\left\vert 0\right\rangle $ and $\left\vert
1\right\rangle $ respectively would be given by Eqs.(\ref{density of spin 0}%
) and (\ref{density of spin 1}) if there were interference effects
between different condensates. In this case, the density
distributions would be quite different from Fig.1-Fig.4. Shown in
Fig.5 would be the density distribution of the wave packets at
time $t=15$ ms with $r=2d$ and $\delta \theta =2\pi /(2k_{M}+1)$.
From Fig.5, we can see that there exists a strong decay and
revival of the density oscillation, and there is even no legible
interference fringe (see Fig.5(b)).

\begin{figure}[h]
\centerline{\includegraphics[width=2cm,angle=-90]{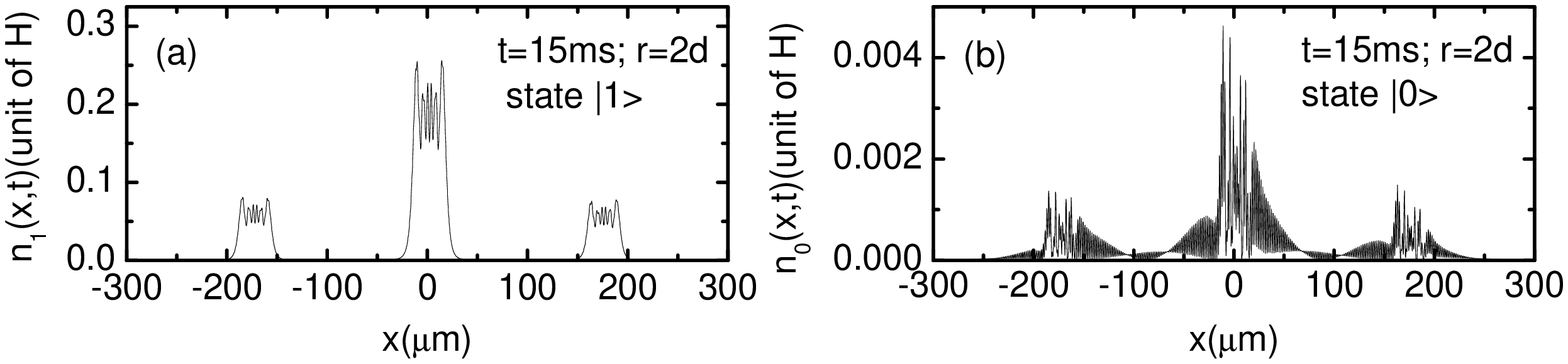}}
\caption{Density distributions in states $\left\vert
1\right\rangle $ (a) and $\left\vert 0\right\rangle $ (b)
respectively provide that there is a fixed phase distribution
among the array of condensates confined in the spin-dependent
optical lattice and there exists interference effects between
different condensates. The time of flight period is $15$ ms. The
phase of
the $k$th condensate is given by $\protect\theta _{k}=2\protect\pi %
(k_{M}+k)/(2k_{M}+1)$ ($k=-k_{M},...,k_{M}$). Here $r=2$ denotes
that initially localized atoms have been delocalized over three
lattice sites.}
\end{figure}

Due to the entanglement of a single atom, however, we predict that
after the nonuniform phase distribution the density distributions
of the wave packets in states $\left\vert 0\right\rangle $ and
$\left\vert 1\right\rangle $ respectively are still given by
Eqs.(\ref{simplified density of spin 0}) and (\ref{simplified
density of spin 1}), i.e., the density distributions will not
change. In other words, for the atoms in the entangled single-atom
state each atom interferes only\ with itself whether the
condensates are full coherent (with fixed relative phases) or
completely independent (with random relative phases), which
implies that the entanglement of a single atom is sufficient for
the interference of BECs in a spin-dependent optical lattice and
the interference effect is irrelevant with the phases of the
individual condensates. This experimental proposal provides a
straight way to test further our theoretical model and prediction.

\section{Discussion and conclusion}

To summarize, we have developed a theoretical model to investigate
the interference of an array of BECs confined in a 1D
spin-dependent optical lattice by calculating the density
distributions and evolution of the atomic wave packets. In such a
system as experienced beforehand a Mott transition and a
spin-dependent transport, each atom can be described by an
entangled single-atom state between the internal (spin) and the
external (spatial wave packet) degrees of freedom . Our
theoretical model is based on an assumption that for the atoms in
the entangled single-atom state each atom interferes only\ with
itself. The results obtained from this model agree well with the
interference patterns observed in a recent experiment \cite{R24},
which in turn verifies the validity of this model and assumption.
In addition, when taking into account the relative phase $\beta $
between the two wave packets of an atom which is obtained during
the transport process, it is found that the symmetry of the
density distributions is broken to a certain extent.

From the present work, it has been shown that due to the
entanglement of a single atom each atom interferes only with
itself (or each condensate interferes only with itself in this
system), i.e., there is no interference effect between two
arbitrary different condensates. In other words, the entanglement
of a single atom is sufficient for the interference of BECs
confined in a spin-dependent optical lattice and the interference
shows no relevancy with the phases of individual condensates.
Finally, an experimental suggestion of nonuniform phase
distribution is proposed to test further our theoretical model and
prediction. The theoretical model presented here can be also
applied to describe the dynamics of BECs trapped in a combined
harmonic and optical lattice potential, wherein the number of
atoms in individual lattice sites is different. Possibly, the
method even can be extended to consider the case of non-perfect
Mott-insulator state.

\begin{acknowledgments}
The authors would like to thank I. Bloch, J. Wang, K. L. Gao, and
Y. Wu for useful discussions. Pertinent comments and suggestions
from the editor and anonymous reviewers are acknowledged. This
work was supported by the National Natural Science Foundation of
China under Grant Nos.10474119,10205011 and 10474117, by the
National Fundamental Research Program of China under Grant
No.001CB309309, and also by funds from the Chinese Academy of
Sciences.
\end{acknowledgments}

\end{document}